                        \newif\ifpaper \newif\ifPDF               %%
                        \newif\ifOUP \newif\ifboyscout            %%
                        \newif\ifdasbuch \newif\ifarticle         %%
                        \newif\ifsolutions

                        \newif\ifboyscout                         %%
                        \boyscouttrue %% commented, WWW/boyscouts %%
                        \newif\ifpreparepdf                       %%
                        \preparepdftrue % hyperlinked pdf default %%

    \boyscoutfalse                 % public, hyperlinked
        \ifboyscout
\documentclass[pre,aps,twocolumn,showpacs,hyperref]{revtex4-1} %or {revtex4-1}
        \else
\documentclass[pre,aps,twocolumn,showpacs,superscriptaddress]{revtex4} %or {revtex4-1}
        \fi

\usepackage[export]{adjustbox} % http://ctan.org/pkg/adjustbox
\graphicspath{{figs/}{../figs/}}  %% directories with graphics

\ifpreparepdf
    \usepackage{color} % dvips allows for colors
    \usepackage[colorlinks]{hyperref} %% hyperlinks
\else % prepare B&W file
\fi
    \usepackage{graphicx}
    \usepackage{ifthen}

\ifpreparepdf % hyperlinked pdf, keep homepage flexible:
    \newcommand{\wwwcb}[1]{
                  {\tt \href{http://ChaosBook.org#1}
              {ChaosBook.org#1}}}

    \newcommand{\arXiv}[1]{
              {\tt \href{http://arXiv.org/abs/#1}{arXiv:#1}}}
\else  %% prepare for postscript printing:
    \newcommand{\wwwcb}[1]{{\tt ChaosBook.org#1}}

    \newcommand{\arXiv}[1]{ {\tt arXiv:#1}}
\fi

        \ifboyscout % if draft, display comments in text
   \newcommand{\PCedit}[1]{{\color{blue}#1}}
   \newcommand{\PC}[2]{\begin{quote}\PCedit{[#1 Predrag] #2}\end{quote}}
   \newcommand{\DLedit}[1]{{\color{red}#1}}
   \newcommand{\DL}[2]{\begin{quote}\DLedit{[#1 Domenico] #2}\end{quote}}
   \definecolor{darkgreen}{rgb}{0,.6,.25}
   \newcommand{\JMHedit}[1]{{\color{darkgreen}#1}}
   \newcommand{\JMH}[2]{\begin{quote}\JMHedit{[#1 Jeffrey] #2}\end{quote}}
        \else  % drop comments
   \newcommand{\PC}[2]{}{}
   \newcommand{\PCedit}[1]{#1}
   \newcommand{\DL}[2]{}{}
   \newcommand{\DLedit}[1]{#1}
   \newcommand{\JMH}[2]{}{}
   \newcommand{\JMHedit}[1]{#1}

        \fi

\newcommand{\rf}     [1] {~\cite{#1}}
\newcommand{\refref} [1] {ref.~\cite{#1}}

\newcommand{\refrefs}[1] {refs.~\cite{#1}}
\newcommand{\refeq}  [1] {(\ref{#1})}
\newcommand{\reffig} [1] {Fig.~\ref{#1}}
\newcommand{\RefFig} [1] {Figure~\ref{#1}}
\newcommand{\refFig} [1] {Fig.~\ref{#1}}
\newcommand{\refsect}[1] {sect.~\ref{#1}}

\newcommand{\beq}{\begin{equation}}
\newcommand{\continue}{\nonumber \\ }

\newcommand{\eeq}{\end{equation}}
\newcommand{\ee}[1] {\label{#1} \end{equation}}
\newcommand{\bea}{\begin{eqnarray}}

\newcommand{\eea}{\end{eqnarray}}

\newcommand{\statesp}{state space}

\newcommand{\eqv}{equi\-lib\-rium}

\newcommand{\eqva}{equi\-lib\-ria}

\newcommand{\po}{periodic orbit}

\newcommand{\dmn}{\ensuremath{\,d}}  %  n-dimensional

\newcommand{\monodromyM}{monodromy matrix} % monodromy matrix, Poincare cut
\newcommand{\optPart}{optimal partition}
\newcommand{\OptPart}{Optimal partition}
\newcommand{\Fokker}{Fokker-Planck}

\newcommand{\ssp}{\ensuremath{x}}    % state space point
\newcommand{\ExpaEig}{\Lambda}
\newcommand{\Lyap}{\ensuremath{\lambda}}            %Lyapunov exponent
\newcommand{\cl}[1]{{n_{#1}}}   % discrete length of a cycle, Predrag
\newcommand{\msr}{{\rho}}               % measure
\newcommand{\pS}{{\cal M}}          % symbol for state space space
\newcommand{\spaceAver}[1]{\left\langle {#1} \right\rangle}
\newcommand{\diffCon}{\ensuremath{D}}       % diffusion constant
\newcommand{\diffTen}{\ensuremath{\Delta}}  % diffusion tensor
\newcommand{\covMat}{\ensuremath{Q}}             % covariance matrix
\newcommand{\Lnoise}[1]{{\cal L}^{#1}}    % noisy evolution operator
\newcommand{\transp}[1]{{#1}{}^\top}
\newcommand{\monodromy}{\ensuremath{M}}   % monodromy matrix, full Poincare cut

\newcommand{\MatrixII}[4]{\left(
\begin{array}{cc}
{#1}  &  {#2} \\
{#3}  &  {#4} \end{array} \right)}

\begin{document}
\title{
Neighborhoods of periodic orbits and
the stationary distribution of a noisy chaotic system
}
\author{Jeffrey M. Heninger}
\affiliation{Department of Physics,  University of Texas, Austin, TX}
\author{Domenico Lippolis\footnote{domenicuzzo@gmail.com - present address:
Faculty of Science, Jiangsu University, Zhenjiang 212013, China.}}
\affiliation{Institute for Advanced Study, Tsinghua University, Beijing, China}
\author{Predrag Cvitanovi\'c}
\affiliation{
                Center for Nonlinear Science and School of Physics,
                Georgia Institute of Technology,
                Atlanta, GA
               }
\date{\today}

	\begin{abstract}
The finest state space resolution that can be achieved in a physical
dynamical system is limited by the presence of noise. In the
weak-noise approximation the stochastic neighborhoods of
deterministic
periodic orbits can be computed from distributions stationary under the
action of a local {\Fokker} operator and its adjoint. We derive explicit
formulae for widths of these distributions in the case of chaotic
dynamics, when the periodic orbits are hyperbolic.
The resulting neighborhoods form a basis for functions on the attractor.
The global stationary distribution, needed for calculation of long-time
expectation values of observables, can be expressed in this basis.
	\end{abstract}

        % What? - JMH

\maketitle
\section{Introduction}

This paper investigates the interplay of deterministic chaotic dynamics
and weak stochastic noise, and proposes a new definition of the
neighborhood of a noisy hyperbolic \statesp\ point.
Such neighborhoods are conjectured to partition the \statesp\ in optimal
way and provide a basis function set for the evaluation of the stationary
distribution.

\subsection{Width of a noisy trajectory}

The basic idea of a stochastic `neighborhood' is that the balance between
the noise broadening of a trajectory  and the deterministic contraction
leads to a probability distribution of finite width,
as opposed to one that spreads with time (diffusion only).
For an orbit that converges to a linearly stable, attractive \eqv, this
neighborhood was computed in 1810 by Laplace\rf{Laplace1810,Jac97} and is
today known as a solution to the Lyapunov equation\rf{Lyapunov1892}, or
the Ornstein-Uhlenbeck process\rf{OrnUhl30}: for a 1-dimensional flow,
the deterministic \eqv\ point is smeared into a Gaussian probability
density centered on it, whose covariance $\covMat=-\diffCon/\Lyap$ is a
balance of the expansion rate $\diffCon$ (diffusion constant) against the
contraction rate $\Lyap<0$. \Fokker\
equation\rf{Risken96} generalizations to higher-dimensional stable \eqva\
and limit cycles (stable \po s) are immediate, provided proper care is
taken of the diffusion along the \po\rf{vKampen92,NaSaWa13}.

What if a \po\ is unstable? Both the diffusion rate and the linearized
stability rate $\Lyap>0$ now expand forward in time, and cannot balance
each other. This problem was solved in \refrefs{LipCvi08,CviLip12} for
repelling \po s with \emph{no contracting directions}, by balancing the
stochastic diffusion against the contraction by the \emph{adjoint}
\Fokker\ operator. The resulting covariance matrix % $\covMat_{ee}$
defines the stochastic neighborhood for a repelling orbit, while the
Ornstein-Uhlenbeck covariance %$\covMat_{cc}$
defines it for a stable orbit. However, neither these stable nor
repelling orbits play a role in chaotic dynamics. The long-time attractors
of chaotic dynamics are organized by an infinity of
\emph{hyperbolic} periodic orbits\rf{gutzwiller71,Ruelle76,DasBuch},
orbits which an ergodic trajectory visits by approaching them along their
stable eigendirections, and leaves along their unstable eigendirections.

The central result of this paper is that techniques developped for
solving the Lyapunov equations\rf{ZhoSalWu99,FaIo01,Varga01} enable us to
define the neighborhood of a \emph{hyperbolic} periodic point by
splitting the covariance matrix $\covMat$ into two (mutually
non-orthogonal) covariance matrices, $\covMat_{cc}$ for contracting
directions, and $\covMat_{ee}$ for the expanding directions.

There are two immediate applications of the notion of the neighborhood of
a hyperbolic point:
(a) `{\optPart}' of the attractor, and
(b) construction of a basis set for the stationary distribution of a
noisy chaotic flow.

\subsection{An \optPart\ from periodic orbits}

While in the idealized deterministic dynamics the \statesp\ can be
resolved arbitrarily finely,  in physical systems noise
always limits the attainable \statesp\ resolution.

This observation had motivated the many limiting resolution estimates for
\statesp\ ‘granularity’ of chaotic systems with background noise. The
idea of an \optPart\ in this context was first introduced in 1983 by
Crutchfield and Packard\rf{CruPack83} who formulated a \statesp\
resolution criterion in terms of a globally averaged ``attainable
information.'' The approach was later generalized and applied to
time-series analysis, where the underlying dynamics is
unknown\rf{DFT03,BuKe05}). A different strategy consists of computing a
transfer matrix between intervals of a uniform grid, and estimating
averages of observables from its eigenvalues and eigenfunctions. First
introduced by Ulam\rf{Ulam60}, this technique has been developed over the
years\rf{Froy-2,Tanner_cars}. All of these approaches (see
\refref{CviLip12} for a review) are based on \emph{global} averages, and
assume that ‘granularity’ is uniform across the \statesp. In contrast,
the main, computationally precise lesson of our work
is that even when the external noise is white, additive,
and globally homogenous, the interplay of noise and nonlinear dynamics
always results in a \emph{local} stochastic neighborhood, whose
covariance depends on both the past and the future noise integrated and
non-linearly convolved with deterministic evolution along the trajectory.
The optimal resolution thus varies from neighborhood to neighborhood, and
has to be computed \emph{locally}.
As was shown in \refref{LipCvi08} for a strictly expanding 1\dmn\ chaotic
map and a given noise, the maximal set of non-overlapping neighborhoods
of \po s can be used to construct an `{\em \optPart}' of the \statesp,
and compute dynamical averages from the associated approximate matrix
\Fokker\ operator.

\subsection{The stationary distribution}
\label{sect:statDistr}

In this paper we utilize our construction of \optPart s to
approximate the {\em stationary probability distribution function} by a
finite sum over Gaussians, one for each neighborhood. When the dynamics
is chaotic, the most one can predict accurately for long times are the
statistical properties of the system, given by the state-space averages
of observables $a(\ssp)$,
\beq
\spaceAver{a} = \int d\ssp \, \msr(\ssp)\, a(\ssp)
\,,
\ee{lt_av_closed}
where the \textit{stationary distribution} (natural
measure\rf{sinai,bowen,ruelle}) $\msr(\ssp)$ is the probability of finding
the system in the state $\ssp$ on the attractor. For a deterministic
system $\msr(\ssp)$ is  a singular, nowhere differentiable distribution
with support on a fractal set, and its numerical computation is usually
not feasible. However, for any physical system the noise washes out fine
details of the dynamics, and the stationary distribution is smooth. Here we
propose a smooth function basis for $\msr(\ssp)$, based on the
{\optPart} of the \statesp.
We develop our formalism for discrete-time dynamical systems and
illustrate it by computing the neighborhoods and estimating $\msr(\ssp)$
for the Lozi map\rf{lozi2}, a simple 2-dimensional discrete-time chaotic
system. The idea is to first partition the attractor into an {\optPart}
set of neighborhoods, and then use the associated local Gaussian
distributions as a finite set of basis functions  for the global
stationary distribution. In the 2-dimensional Lozi example, our estimates
for the stationary distribution are consistent with those obtained by the
direct numerical estimation of the lattice-discretized probability
densities computed from long stochastic (Langevin) trajectories.

\section{The neighborhood of a hyperbolic point}

An autonomous discrete-time stochastic dynamical system $(\pS, f,
\diffTen)$ can be defined by specifying a \statesp\  $\pS$, a
deterministic map $f: \pS \rightarrow \pS$, and an additive noise
covariance matrix (diffusion tensor) $\diffTen = \diffTen(\ssp)$. In one
time step, an initial Dirac-delta density distribution $\msr_a(\ssp)$
located at $x$ is smeared out into a Gaussian ellipsoid $\msr_{a+1}(y)$
centered at
$y = f(\ssp)$, with covariance $\diffTen(\ssp)$. This defines the kernel
of the {\Fokker} evolution operator in $d$ dimensions\rf{Risken96}
\bea
\Lnoise{}_{FP}(y,x)\, dx
    &=&
e^{-\frac{1}{2} \transp{(y-f(\ssp))} \frac{1}{\diffTen(\ssp)} (y-f(\ssp))}
\, [dx]
\continue
{[dx]} &=& d^dx/\det(2 \pi\diffTen)^{1/2}
\,.
\label{NoiseTimeStep}
\eea
Consider a trajectory
\(
\{\ssp_a\} =
(\ssp_a,\ssp_{a+1},\ssp_{a+2},\cdots
)
\)
generated by the deterministic evolution rule $\ssp_{a+1} = f(\ssp_a)$, and
shift the coordinates in each $\ssp_a$ neighborhood to $\ssp = \ssp_a + z_a$. In the
vicinity of $\ssp_a$ the dynamics can be linearized as
\(
z_{a+1} = \monodromy_a z_a\,,
\) %\ee{LinearizedMap}
where $\monodromy_a = \partial f(\ssp_a)$ is the one time-step Jacobian
matrix.

Prepare the initial
density of trajectories $\msr_a(z_a)$ in the $\ssp_a$ neighborhood
as a normalized Gaussian distribution $\msr(z_a,\covMat_a)$,
centered at $z_a=0$, with a strictly positive-definite covariance matrix
$\covMat_a$. The support of density $\msr(z_a,\covMat_a)$ can be
visualized as an ellipsoid with axes oriented along the eigenvectors of
$\covMat_a$.
The linearized {\Fokker} operator
\[ %beq
\Lnoise{}(z_{a+1},z_a)\,dz_a  =
e^{-\transp{(z_{a+1}-M_a z_a)}\frac{1}{2\Delta_a}(z_{a+1}-M_a z_a)}
\,[dz_a]
\] %ee{locLnoise}
maps this distribution one step forward in time into another Gaussian
\beq
\msr(z_{a+1},\covMat_{a+1})
\,=\,
\int dz_a \, \Lnoise{}{}(z_{a+1},z_a)\, \msr(z_a,\covMat_a)
\,,
\ee{JMHFokPla}
with the covariance matrix deformed by the dynamics and spread out by the
noise, as given by the discrete Lyapunov equation\rf{Lyapunov1892,GaQu95},
\beq
\covMat_{a+1} = \monodromy_a \covMat_a \transp{\monodromy_a} + \diffTen_a
\,.
\ee{FokPla4Gauss}
In other words, the two covariance matrices, (i) the deterministically
transported $\covMat_a \to \monodromy_a \covMat_a \transp{\monodromy_a}$,
and (ii) the noise diffusion tensor $\diffTen_a$, add together in the
usual manner, as squares of errors.

Similarly,
density evolution for dynamics with strictly expanding Jacobian matrices
$\monodromy_a$ can be described by the action of the adjoint
{\Fokker} operator\rf{LipCvi08,CviLip12}, with kernel
\[ %\beq
\Lnoise{\dagger}(y,x)\, dy =
e^{-\frac{1}{2} \transp{(y-f(\ssp))} \frac{1}{\diffTen(\ssp)} (y-f(\ssp))}
\, [dy]
\,.
\]
The adjoint {\Fokker} operator expresses the current density
$\msr_{a}$ as the convolution of its image $\msr_{a+1}$ with the noisy
dynamics
\[
\msr_a(z_a,\covMat_{a}) =
\int dz_{a+1} \Lnoise{\dagger}{}(z_a,z_{a+1})\msr_{a+1}(z_{a+1},\covMat_{a+1})
%\propto e^{-\frac{1}{2} \transp{z_a} \frac{1}{\covMat_a} z_a}
\,.
\]
%\ee{adjevol}
Like in the forward evolution, we may substitute a Gaussian density into
this equation
to obtain the discrete adjoint Lyapunov equation for the
covariance matrices
\beq
\monodromy_a \covMat_a \transp{\monodromy_a}
  \,=\, \covMat_{a+1} + \diffTen_{a}
\,. \label{BackwardEvolution}
\eeq
We show in what follows that, if the Jacobian matrices $\monodromy_a$
have all eigenvalues strictly contracting (expanding), any initial
Gaussian converges to an invariant density under the action of the
(adjoint) {\Fokker} operator. Consider first the case of a map $f(\ssp)$
with a stable fixed point at $\ssp_a$ (at $z = z_a = 0$). The covariance
matrix transforms as
\bea
\covMat &=& \diffTen + \monodromy \diffTen \transp{\monodromy}
  + \monodromy^2 \diffTen (\transp{\monodromy})^2 + \cdots \nonumber\\
&=& \sum\limits_{m,n=0}^\infty
\monodromy^n \diffTen (\transp{\monodromy})^m \delta_{mn}
\,.
\label{JMH:cov_sum}
\eea
By inserting the Fourier representation of
Kronecker $\delta_{mn}$ into \refeq{JMH:cov_sum}, we
can recast this expression into the resolvent form
\bea
\covMat &=& \int\nolimits_0^{2\pi} \frac{d\theta}{2\pi} \sum\limits_{m,n=0}^{\infty}
(e^{-i\theta} \monodromy)^n \diffTen (e^{i\theta} \transp{\monodromy})^m \nonumber\\
&=& \int\nolimits_0^{2\pi} \frac{d\theta}{2\pi} \frac{1}{{1} - e^{-i\theta} \monodromy}
\diffTen \frac{1}{{1} - e^{i\theta} \transp{\monodromy}}.
\label{IntegralCovMat}
\eea
We do the same in the expanding case, by using the adjoint evolution
\bea
\monodromy \covMat \transp{\monodromy} &=&
\sum\limits_{m,n=0}^\infty
  \frac{1}{\monodromy^n} \diffTen \frac{1}{(\transp{\monodromy})^m} \delta_{nm}
\nonumber \\
&=& \int\nolimits_0^{2\pi} \frac{d\theta}{2\pi} \sum\limits_{m,n=0}^{\infty}
\left( \frac{e^{i\theta}}{\monodromy} \right)^n \diffTen \left( \frac{e^{-i\theta}} {\transp{\monodromy}} \right)^m \nonumber\\
&=& \int\nolimits_0^{2\pi} \frac{d\theta}{2\pi} \frac{1}{{1} - e^{i\theta} /{\monodromy}} \diffTen
\frac{1}{{1} - e^{-i\theta} /{\transp{\monodromy}}},
\eea
which is then easily reduced to~\refeq{IntegralCovMat}, so that
the resolvent form is the same regardless of whether $\monodromy$ is
expanding or contracting.
This result comes particularly handy when we deal with a
hyperbolic fixed point, that is
when $\monodromy_a$ has both expanding and contracting
eigenvalues.
The {\monodromyM} is not symmetric, and it cannot be diagonalized by an
orthogonal transformation, but its expanding and the contracting parts can
be separated with a similarity transformation $S$ that brings
$\monodromy$ to a block-diagonal form,
\beq
\ExpaEig \equiv S^{-1} \monodromy S =
\MatrixII{\ExpaEig_e}{0}{0}{\ExpaEig_c}
\,.
\ee{JHMdiagonalizeM}
Here the blocks $\ExpaEig_e$ and $\ExpaEig_c$ contain all expanding,
contracting eigenvalues of the {\monodromyM}, respectively.
The covariance matrix $\covMat =S\hat{\covMat}\transp{S}$ is not
block-diagonalized by the above similarity transformation, but consider
the four blocks
\beq
\hat{\covMat}_{ij} =
\int\nolimits_0^{2\pi} \frac{d\theta}{2\pi} \frac{1}{{1} - e^{-i\theta}\ExpaEig_i}
\hat{\diffTen}_{ij} \frac{1}{{1} - e^{i\theta} \transp{\ExpaEig}_j}
\,,
\ee{block_res}
where $\hat{\diffTen} \equiv S^{-1} \diffTen \transp{(S^{-1})}$, and $i,j
\in \{c,e\}$, where $\{c,e\}$ denotes $\{$contracting, expanding$\}$.
This expression may be evaluated as
a contour integral around the unit circle in the complex
plane\rf{ZhoSalWu99,Varga01}
\beq
\hat{\covMat}_{ij} =
\frac{1}{2\pi i} \oint\nolimits_\Gamma dz \frac{1}{z {1} - \ExpaEig_i}\hat{\diffTen}_{ij}
\frac{1}{{1} - z \transp{\ExpaEig}_j}
\,.
\eeq
The diagonal blocks
$\hat{\covMat}_{cc}$, $\hat{\covMat}_{ee}$
have either all expanding or all contracting eigenvalues,
meaning at least one pole inside and one pole outside the unit
circle, and the residue theorem yields a non-vanishing result for the integral.
Consider next the off-diagonal block $\hat{\covMat}_{ce}$ with$\Lambda_i$
contracting and $\Lambda_j$ expanding:
in this case the poles all lie outside the unit circle, and the integral
vanishes. The remaining off-diagonal block having $\Lambda_i$ expanding and $\Lambda_j$
contracting must also vanish when integrated, due to the symmetry of $\hat{\covMat}$,
which is therefore block-diagonal,
\beq
\covMat = S \MatrixII{\hat{\covMat}_{ee}}{0}{0}{\hat{\covMat}_{cc}} \transp{S}
\,.
\ee{covMat_tsf}

These results are easily extended to a periodic orbit $p$ of period
$\cl{p}$, since any point $\ssp_a$ of the orbit is a fixed point of the
$\cl{p}$th iterate of the map.
The forward and adjoint evolution equations~\refeq{FokPla4Gauss}
and~\refeq{BackwardEvolution} for the covariance matrix,
as well as the resolvent~\refeq{IntegralCovMat} all
still hold, with some changes in the notation:
each periodic point $\ssp_a$ has its own neighborhood, with its own
covariance matrix $Q_a$. The \monodromyM\ $M_a$ of $\ssp_a$ now evolves
$n_p$ steps along the orbit
\[
\monodromy_a^{n_p} =
\monodromy_{a+n_p-1} \cdots \monodromy_{a+2} \monodromy_{a+1} \monodromy_a,
\]
while the diffusion tensor $\Delta_a$
now accounts for the total noise accumulated along the periodic orbit,
\bea \label{diffPerOrb} \diffTen_{p,a} &\equiv&
 \sum\limits_{i=0}^{\cl{p}-1} \ \monodromy_{a+i+1}^{n_p-i-1} \ \diffTen_{a+i}
\ \transp{\monodromy_{a+i+1}^{n_p-i-1}}
\,.
\eea

\section{\OptPart\ and stationary distribution}

\begin{figure}
	(a) \includegraphics[width=.2\textwidth,valign=t]{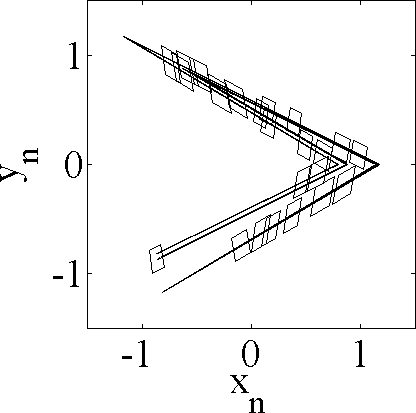}
	(b) \includegraphics[width=.2\textwidth,valign=t]{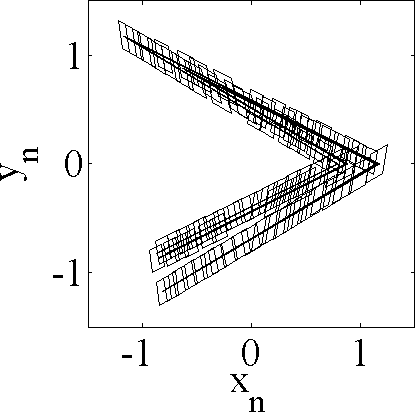} \\
	(c)	\includegraphics[width=.42\textwidth,valign=t]{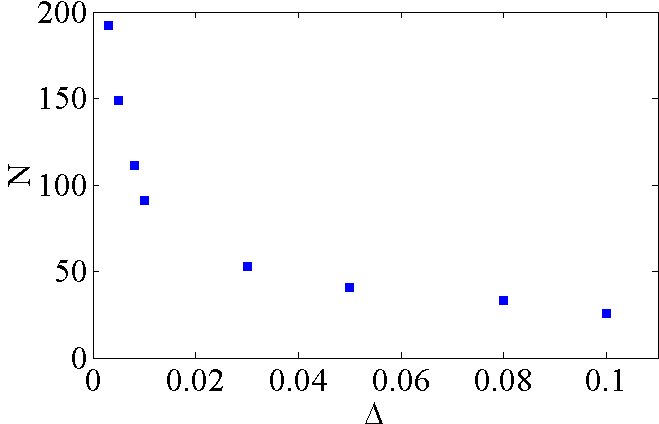}
	\caption{\label{Partition}
    Building the partition for the Lozi attractor, for an isotropic
    constant diffusion tensor $\delta_{ij}\Delta$.
    In frames (a) and (b) $\Delta = 0.01$.
	The deterministic attractor is the fractal structure in the
    background of each picture. Stochastic neighborhoods of a set of
    periodic points are indicated by their standard deviation
    parallelograms.
	(a) An initial partition: with only periodic points of periods
        $\leq 5$ much of the attractor remains to be covered.
	(b) The final, \optPart\ covers the whole attractor, with no
        pair of neighborhoods overlapping by more than 50\%.
	(c) $N$,
        the number of neighborhoods needed to achieve the \optPart\ for
        a given noise strength $\Delta$.
    }
\end{figure}

At this point our strategy is to build a partition out of
\textit{neighborhoods} of the periodic points, each defined
by means of the stationarity condition~\refeq{IntegralCovMat}:
solve for the expanding and contracting blocks of~\refeq{covMat_tsf}
separately, and draw a parallelogram on the supports of the resulting Gaussians,
with axes oriented along the eigenvectors of the covariance matrices $\covMat_{ee}$
and $\covMat_{cc}$, their widths
given by one standard deviation along each direction. We say that two
neighborhoods overlap if they do so by at least $50\%$ of their areas (consistent
with the $1\sigma$ confidence interval chosen as overlapping threshold in
\refref{LipCvi08}).
For a typical chaotic map, periodic points are dense in the
deterministic
attractor\rf{devnmap}, which we now aim to cover entirely with the minimum number of
neighborhoods possible.
We do so via the following algorithm:
\textit{i)} Find periodic points of period $n_p=1,2,...$, and their corresponding neighborhoods.
\textit{ii)} If any neighborhood overlaps with the neighborhood of a shorter periodic point,
then it is discarded
and the neighborhood of lower period occupying the same area is instead kept in the partition.
\textit{iii)} Among groups of neighborhoods of the same period,
discard those that overlap, while keep
the rest in the partition.
\textit{iv)} The algorithm stops
when the attractor is fully covered and no further non-overlapping
neighborhoods can be found. An example is shown in \reffig{Partition} for
the two-dimensional Lozi attractor\rf{lozi2}.
%    \PC{2015-10-25}{Replace `Number of Neighborhoods' in
%                \reffig{Partition}\,(c) by $N$}

The main utility of a good partition is that it provides a basis for an
accurate and efficient estimate of long-time averages of observables
defined on the dynamical system, of the form~\refeq{lt_av_closed}.
As explained in \refsect{sect:statDistr}, our goal here is to determine
the stationary distribution $\msr(\ssp)$. For that purpose, we use as
basis Gaussian ellipsoids that satisfy the \textit{local} stationarity
condition~\refeq{IntegralCovMat} in each neighborhood of the \optPart.
A set of Gaussians centered at every point in the \statesp\ $\pS$ forms an
overcomplete, non-orthogonal basis for functions in $L^2(\pS)$, as is
well known from the study of coherent states of quantum harmonic
oscillators\rf{GottYan}.
Our (also overcomplete and non-orthogonal) set
of Gaussians is centered only on periodic points, which are dense in the
deterministic attractor, but not in the entire \statesp.
Therefore our basis is designed to resolve the structure of any function
with support on the hyperbolic `strange set' (an attractor or a
repeller). The Gaussians are constructed so that their widths balance the
noise spreading and the (time-forward or -backward) contraction of the
deterministic dynamics.
In the transverse directions, the basis gives the width of the global
stationary distribution, locally everywhere determined by the balance
between noise and dynamics. Along the attractor, the basis determines the
minimum number of neighborhoods needed to fully resolve the structure of
the stationary distribution.

There are numerical methods (such as refinements of Ulam's
method\rf{ErmShep12}) that identify the asymptotic attractor by
running long noisy trajectories, dropping the transients, and covering
the attractor so revealed by a finite number of boxes. These algorithms
have no a priori information about how the stationary distribution
behaves transversely to the deterministic attractor, and they may easily
overestimate the number of basis elements needed to resolve this
structure. In contrast, in our approach the transverse structure is
automatically accounted for by the local balance between the noise and
the deterministic contraction along the stable, transverse directions,
given by covariance matrix block $\hat{\covMat}_{cc}$ in
\refeq{covMat_tsf}. Furthermore, estimating $\msr(\ssp)$ by binning long
noisy trajectory over a finite number of attractor-covering boxes is
feasible only in a low-dimensional \statesp, while \refeq{covMat_tsf} can
be computed for \statesp\ of any dimension.

In discrete time dynamics, the stationary distribution is the
ground-state eigenfunction of the Fokker-Planck evolution
operator \refeq{NoiseTimeStep} with escape rate $\gamma$,
\beq
	\Lnoise{} \, \msr(\ssp) = e^{-\gamma}\msr(\ssp)  \,.
\ee{StationCond}
In order to estimate the stationary distribution, we write it as a sum
over the neighborhoods of the periodic points:
\beq
\msr(\ssp) = \sum\limits_{a=1}^N h_a \, \phi_a(\ssp),
\label{rho_expn}
\eeq
where $\phi_a=e^{-\transp{\ssp}_aQ_a\ssp_a}$ are the Gaussian basis
functions, with  $Q_a$ given by~\refeq{covMat_tsf}, and the coefficients
$\{h_a\}$ to be determined. The truncation of the
expansion~\refeq{rho_expn} to $N$ basis functions follow from our
\optPart. We estimate the coefficients $h_a$ by minimizing the cost
function
\beq
    \int \left( \sum\limits_{a=1}^N h_a (\Lnoise{}
            - e^{-\gamma}) \phi_a(\ssp) \right)^2 dx,
\ee{Error}
together with the normalization constraint for $\msr(\ssp)$. We can also
estimate the escape rate of the system by minimizing the error with
respect to $e^{-\gamma}$.
\begin{figure}
	(a) \includegraphics[width=.20\textwidth,valign=t]{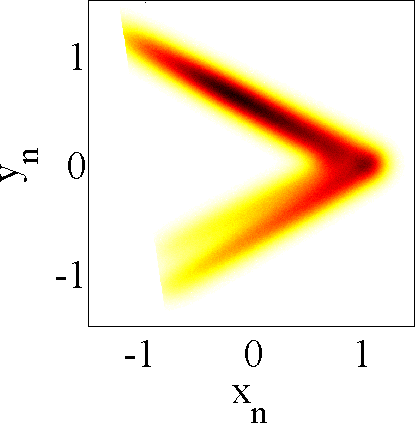}
	(b) \includegraphics[width=.20\textwidth,valign=t]{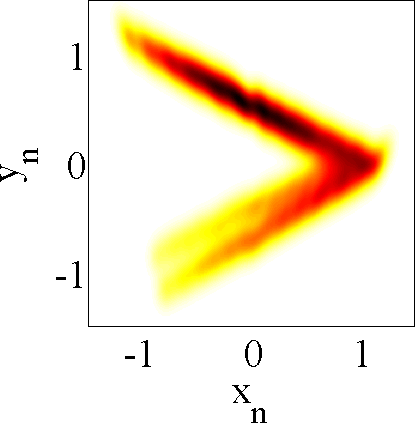}
	\caption{\label{BrutusD01}
(Color online)
The stationary distribution of the Lozi map with $\diffTen= 0.01$. (a) A
direct numerical calculation obtained by running noisy trajectories for a
long time.
(b) The stationary distribution calculated with the \optPart\  method.
    }
\end{figure}

\begin{figure}
	(a) \includegraphics[width=.45\textwidth,valign=t]{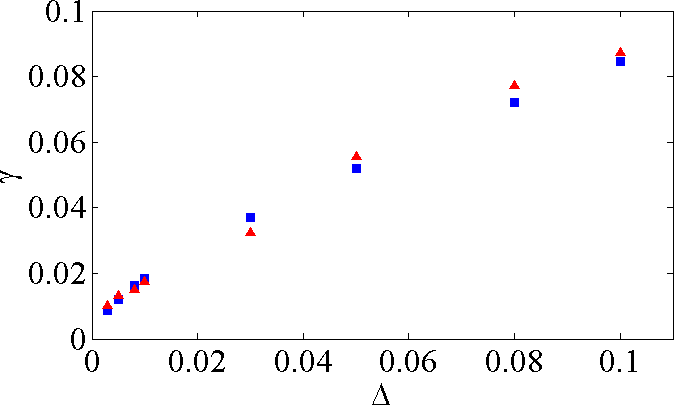} \\
	(b) \includegraphics[width=.45\textwidth,valign=t]{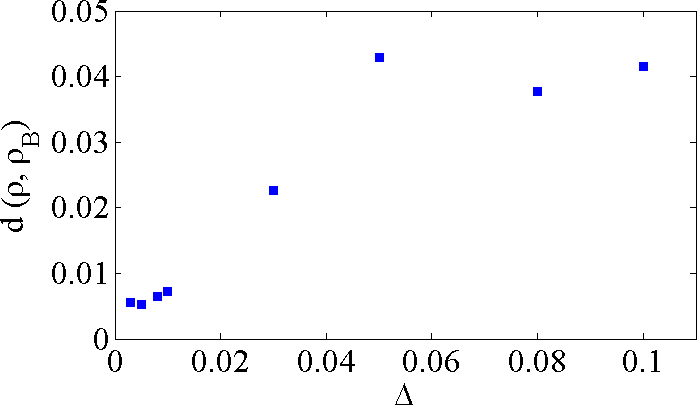}
	\caption{\label{Results}
	(a) The escape rate from the attractor as the function of the
        noise strength $\diffTen$.
          Squares: uniform grid discretization method.
          Triangles: \optPart.
	(b) The normalized $L^2$ distance $d(\msr,\msr_B)$ between the
        corresponding stationary distributions.
    %	For all values of the noise, the two distributions are within 5\% of each other.
	}
\end{figure}

As an example, we apply the procedure to the Lozi map\rf{lozi2}
\bea
\nonumber
x_{n+1} &=& 1- a|x_n| + b \ y_n \\
y_{n+1} &=& x_n
\label{lozi_map}
\eea
with parameters $a = 1.85, b = 0.3$ and isotropic, constant diffusion
tensor $\delta_{ij}\Delta$ with $\Delta$ ranging in the interval $[0.003,0.1]$.
\RefFig{Partition}\,(c), which shows the number $N$ of neighborhoods
required by the \optPart\ for a given $\Delta$, illustrates the
efficiency of our method: $N$ goes from tens to few hundreds in the noise
range considered. In order to test our algorithm, we also estimate
$\msr(\ssp)$ and $\gamma$ by a direct numerical simulation.
The $(x,y)$ \statesp\ is divided into uniform mesh
$6.4\times 10^5$ bins; we follow long stochastic trajectories and count how
many times they visit each bin.
The stationary distribution $\msr_B(\ssp)$ is then the normalized
frequency distribution of the whole grid. The deterministic Lozi map has
a fixed point at the edge of attractor, whose stable manifold is the
boundary of the deterministic basin of attraction. The noise makes it possible for a
stochastic trajectory to cross this boundary and escape. We compute the escape rate
as the ratio of the total number of points in the noisy trajectories to
the number of escapes. \refFig{BrutusD01} shows an example of the
stationary distribution estimated with both methods, while in
\reffig{Results} we compare quantitatively the two procedures. In
particular, we estimate the relative error between the stationary
distribution $\msr$ computed with the \optPart\ and $\msr_{B}$ computed
on the uniform grid, by using a normalized $L^2$ distance, as
\beq
 d(\msr,\msr_B) =	\frac{\int \left( \msr(\ssp) - \msr_{B}(\ssp) \right)^2 dx}
	{\int \left( \msr_{B}(\ssp) \right)^2 d\ssp}.
\ee{L2Diff}
The two distributions are within $5\%$ of each other, whereas the escape
rates differ by at most 10\%, over a range of $\Delta$ that spans
two orders of magnitude.

\section{Discussion}

In conclusion, we have generalized the \optPart\ hypothesis first
formulated in\rf{LipCvi08} to hyperbolic maps in arbitrary dimension, and
tested the method on a two-dimensional system with weak white isotropic
noise. As noise induces a finite resolution of the \statesp\ of any
physical system, finite numbers of neighborhoods suffice to partition the
\statesp\ explored by chaotic dynamics, and to estimate long-time
averages of observables. Here we have used the deterministic unstable
periodic orbits as the skeleton on which to build an \optPart\ for the
noisy \statesp. First we determine a local stationary distribution in
the neighborhood of each periodic point by balancing the noise against
the deterministic expansion or contraction. From the separation of
expanding and contracting blocks in the covariance matrix that
characterizes the Gaussian approximation to the local stationary
distribution, we carve out a precise definition of neighborhood, the
constituent of our partition, which is then used to approximate the
global stationary distribution, estimate the escape rate (for open systems that
allow escape), and any long-time
averaged observable. Numerical tests confirm that the accuracy of our
method is comparable to that of a
uniform grid discretization, but the number neighborhoods required for
our \optPart\ ($\sim10$ to 100) is three-four orders of magnitude
smaller than the number of bins used in the uniform grid discretization
method ($\sim10^5$).

The problems dynamical chaos (or `turbulence') theory faces nowadays are
not two- but high-, even infinite-dimensional. Today it is possible to
compute numerically exact periodic orbits (`recurrent
flows'\rf{focusPOT}) in a variety of physically realistic turbulent fluid
flows\rf{N90,GHCW07}, but these calculations are at the limit of what
current codes can do, and we hope that the methods presented here can
provide sharp criteria for when a sufficient number of such solutions has
been computed. Furthermore, unlike the
uniform grid discretization,
our partitions are smart, since they rely on the periodic orbits of the
deterministic system as skeleton of the dynamics, as well as efficient,
due to the finite (and optimal!) numbers of neighborhoods and
corresponding basis functions.
This, we believe, should make our algorithm  less costly to implement
than direct numerical simulations in higher dimensions, where
discretizations would be impractical. With some modifications and
application of Poincar\'e sections, the formalism can be applied to
continuous time flows as well\rf{gasp02,CviLip12}. Outstanding challenges
include dealing with the lack of hyperbolicity in higher dimensions
(marginal directions were treated in \refref{CviLip12} for $1d$ maps), as
well as extending the definition of neighborhood to other time-invariant
sets, such as relative periodic orbits and partially hyperbolic invariant
manifolds.
Further technical issues, such as improving the efficiency of the
minimization algorithm by modifying the basis of functions used in the
computation of the stationary distribution, are also part of our agenda.

\begin{acknowledgments}
We are indebted to P. M. Svetlichnyy for suggesting that we use error
minimization to find the global stationary distribution.
JMH\
thanks Presidential Undergraduate Research Award for partial support,
and DL for the hospitality at IASTU, Beijing.
PC\ thanks the family of late G.~Robinson,~Jr..
and
NSF~DMS-1211827 for  partial support.
DL\ acknowledges support from the National Science Foundation of
China (NSFC), International Young Scientists (11450110057-041323001).
\end{acknowledgments}

 \bibliography{../bibtex/lippolis}

\begin{thebibliography}{33}%
\makeatletter
\providecommand \@ifxundefined [1]{%
 \@ifx{#1\undefined}
}%
\providecommand \@ifnum [1]{%
 \ifnum #1\expandafter \@firstoftwo
 \else \expandafter \@secondoftwo
 \fi
}%
\providecommand \@ifx [1]{%
 \ifx #1\expandafter \@firstoftwo
 \else \expandafter \@secondoftwo
 \fi
}%
\providecommand \natexlab [1]{#1}%
\providecommand \enquote  [1]{``#1''}%
\providecommand \bibnamefont  [1]{#1}%
\providecommand \bibfnamefont [1]{#1}%
\providecommand \citenamefont [1]{#1}%
\providecommand \href@noop [0]{\@secondoftwo}%
\providecommand \href [0]{\begingroup \@sanitize@url \@href}%
\providecommand \@href[1]{\@@startlink{#1}\@@href}%
\providecommand \@@href[1]{\endgroup#1\@@endlink}%
\providecommand \@sanitize@url [0]{\catcode `\\12\catcode `\$12\catcode
  `\&12\catcode `\#12\catcode `\^12\catcode `\_12\catcode `\%12\relax}%
\providecommand \@@startlink[1]{}%
\providecommand \@@endlink[0]{}%
\providecommand \url  [0]{\begingroup\@sanitize@url \@url }%
\providecommand \@url [1]{\endgroup\@href {#1}{\urlprefix }}%
\providecommand \urlprefix  [0]{URL }%
\providecommand \Eprint [0]{\href }%
\providecommand \doibase [0]{http://dx.doi.org/}%
\providecommand \selectlanguage [0]{\@gobble}%
\providecommand \bibinfo  [0]{\@secondoftwo}%
\providecommand \bibfield  [0]{\@secondoftwo}%
\providecommand \translation [1]{[#1]}%
\providecommand \BibitemOpen [0]{}%
\providecommand \bibitemStop [0]{}%
\providecommand \bibitemNoStop [0]{.\EOS\space}%
\providecommand \EOS [0]{\spacefactor3000\relax}%
\providecommand \BibitemShut  [1]{\csname bibitem#1\endcsname}%
\let\auto@bib@innerbib\@empty
%</preamble>
\bibitem [{\citenamefont {Laplace}(1810)}]{Laplace1810}%
  \BibitemOpen
  \bibfield  {author} {\bibinfo {author} {\bibfnamefont {P.~S.}\ \bibnamefont
  {Laplace}},\ }\href@noop {} {\bibfield  {journal} {\bibinfo  {journal} {Mem.
  Acad. Sci. (I), XI, Section V.}\ ,\ \bibinfo {pages} {375}} (\bibinfo {year}
  {1810})}\BibitemShut {NoStop}%
\bibitem [{\citenamefont {Jacobsen}(1997)}]{Jac97}%
  \BibitemOpen
  \bibfield  {author} {\bibinfo {author} {\bibfnamefont {M.}~\bibnamefont
  {Jacobsen}},\ }\href@noop {} {\bibfield  {journal} {\bibinfo  {journal}
  {Bernoulli}\ }\textbf {\bibinfo {volume} {2}},\ \bibinfo {pages} {271}
  (\bibinfo {year} {1997})}\BibitemShut {NoStop}%
\bibitem [{\citenamefont {Lyapunov}(1992)}]{Lyapunov1892}%
  \BibitemOpen
  \bibfield  {author} {\bibinfo {author} {\bibfnamefont {A.~M.}\ \bibnamefont
  {Lyapunov}},\ }\href@noop {} {\bibfield  {journal} {\bibinfo  {journal} {Int.
  J. Control}\ }\textbf {\bibinfo {volume} {55}},\ \bibinfo {pages} {531}
  (\bibinfo {year} {1992})}\BibitemShut {NoStop}%
\bibitem [{\citenamefont {{Uhlenbeck}}\ and\ \citenamefont
  {{Ornstein}}(1930)}]{OrnUhl30}%
  \BibitemOpen
  \bibfield  {author} {\bibinfo {author} {\bibfnamefont {G.~E.}\ \bibnamefont
  {{Uhlenbeck}}}\ and\ \bibinfo {author} {\bibfnamefont {L.~S.}\ \bibnamefont
  {{Ornstein}}},\ }\href@noop {} {\bibfield  {journal} {\bibinfo  {journal}
  {Phys. Rev.}\ }\textbf {\bibinfo {volume} {36}},\ \bibinfo {pages} {823}
  (\bibinfo {year} {1930})}\BibitemShut {NoStop}%
\bibitem [{\citenamefont {Risken}(1996)}]{Risken96}%
  \BibitemOpen
  \bibfield  {author} {\bibinfo {author} {\bibfnamefont {H.}~\bibnamefont
  {Risken}},\ }\href@noop {} {\emph {\bibinfo {title} {The {Fokker-Planck}
  Equation}}}\ (\bibinfo  {publisher} {Springer},\ \bibinfo {address} {New
  York},\ \bibinfo {year} {1996})\BibitemShut {NoStop}%
\bibitem [{\citenamefont {van Kampen}(1992)}]{vKampen92}%
  \BibitemOpen
  \bibfield  {author} {\bibinfo {author} {\bibfnamefont {N.~G.}\ \bibnamefont
  {van Kampen}},\ }\href@noop {} {\emph {\bibinfo {title} {Stochastic Processes
  in Physics and Chemistry}}}\ (\bibinfo  {publisher} {North-Holland},\
  \bibinfo {address} {Amsterdam},\ \bibinfo {year} {1992})\BibitemShut
  {NoStop}%
\bibitem [{\citenamefont {Nakanishi}\ \emph {et~al.}(2013)\citenamefont
  {Nakanishi}, \citenamefont {Sakaue},\ and\ \citenamefont {Wakou}}]{NaSaWa13}%
  \BibitemOpen
  \bibfield  {author} {\bibinfo {author} {\bibfnamefont {H.}~\bibnamefont
  {Nakanishi}}, \bibinfo {author} {\bibfnamefont {T.}~\bibnamefont {Sakaue}}, \
  and\ \bibinfo {author} {\bibfnamefont {J.}~\bibnamefont {Wakou}},\ }\href
  {\doibase 10.1063/1.4834636} {\bibfield  {journal} {\bibinfo  {journal} {J.
  Chem. Phys.}\ }\textbf {\bibinfo {volume} {139}},\ \bibinfo {pages} {214105}
  (\bibinfo {year} {2013})}\BibitemShut {NoStop}%
\bibitem [{\citenamefont {Lippolis}\ and\ \citenamefont
  {Cvitanovi\'c}(2010)}]{LipCvi08}%
  \BibitemOpen
  \bibfield  {author} {\bibinfo {author} {\bibfnamefont {D.}~\bibnamefont
  {Lippolis}}\ and\ \bibinfo {author} {\bibfnamefont {P.}~\bibnamefont
  {Cvitanovi\'c}},\ }\href@noop {} {\bibfield  {journal} {\bibinfo  {journal}
  {Phys. Rev. Lett.}\ }\textbf {\bibinfo {volume} {104}},\ \bibinfo {pages}
  {014101} (\bibinfo {year} {2010})},\ \bibinfo {note}
  {\arXiv{0902.4269}}\BibitemShut {NoStop}%
\bibitem [{\citenamefont {Cvitanovi\'c}\ and\ \citenamefont
  {Lippolis}(2012)}]{CviLip12}%
  \BibitemOpen
  \bibfield  {author} {\bibinfo {author} {\bibfnamefont {P.}~\bibnamefont
  {Cvitanovi\'c}}\ and\ \bibinfo {author} {\bibfnamefont {D.}~\bibnamefont
  {Lippolis}},\ }in\ \href {\doibase 10.1063/1.4745574} {\emph {\bibinfo
  {booktitle} {Let's Face Chaos through Nonlinear Dynamics}}},\ \bibinfo
  {editor} {edited by\ \bibinfo {editor} {\bibfnamefont {M.}~\bibnamefont
  {Robnik}}\ and\ \bibinfo {editor} {\bibfnamefont {V.~G.}\ \bibnamefont
  {Romanovski}}}\ (\bibinfo  {publisher} {Am. Inst. of Phys.},\ \bibinfo
  {address} {Melville, New York},\ \bibinfo {year} {2012})\ pp.\ \bibinfo
  {pages} {82--126},\ \bibinfo {note} {\arXiv{1206.5506}}\BibitemShut {NoStop}%
\bibitem [{\citenamefont {Gutzwiller}(1971)}]{gutzwiller71}%
  \BibitemOpen
  \bibfield  {author} {\bibinfo {author} {\bibfnamefont {M.~C.}\ \bibnamefont
  {Gutzwiller}},\ }\href@noop {} {\bibfield  {journal} {\bibinfo  {journal} {J.
  Math. Phys.}\ }\textbf {\bibinfo {volume} {12}},\ \bibinfo {pages} {343}
  (\bibinfo {year} {1971})}\BibitemShut {NoStop}%
\bibitem [{\citenamefont {Ruelle}(1976)}]{Ruelle76}%
  \BibitemOpen
  \bibfield  {author} {\bibinfo {author} {\bibfnamefont {D.}~\bibnamefont
  {Ruelle}},\ }\href@noop {} {\bibfield  {journal} {\bibinfo  {journal}
  {Invent. Math.}\ }\textbf {\bibinfo {volume} {34}},\ \bibinfo {pages} {231}
  (\bibinfo {year} {1976})}\BibitemShut {NoStop}%
\bibitem [{\citenamefont {Cvitanovi\'{c}}\ \emph {et~al.}(2015)\citenamefont
  {Cvitanovi\'{c}}, \citenamefont {Artuso}, \citenamefont {Mainieri},
  \citenamefont {Tanner},\ and\ \citenamefont {Vattay}}]{DasBuch}%
  \BibitemOpen
  \bibfield  {author} {\bibinfo {author} {\bibfnamefont {P.}~\bibnamefont
  {Cvitanovi\'{c}}}, \bibinfo {author} {\bibfnamefont {R.}~\bibnamefont
  {Artuso}}, \bibinfo {author} {\bibfnamefont {R.}~\bibnamefont {Mainieri}},
  \bibinfo {author} {\bibfnamefont {G.}~\bibnamefont {Tanner}}, \ and\ \bibinfo
  {author} {\bibfnamefont {G.}~\bibnamefont {Vattay}},\ }\href@noop {} {\emph
  {\bibinfo {title} {Chaos: Classical and Quantum}}}\ (\bibinfo  {publisher}
  {Niels Bohr Institute},\ \bibinfo {address} {Copenhagen},\ \bibinfo {year}
  {2015})\ \bibinfo {note} {{\wwwcb{}}}\BibitemShut {NoStop}%
\bibitem [{\citenamefont {Zhou}\ \emph {et~al.}(1999)\citenamefont {Zhou},
  \citenamefont {Salomon},\ and\ \citenamefont {Wu}}]{ZhoSalWu99}%
  \BibitemOpen
  \bibfield  {author} {\bibinfo {author} {\bibfnamefont {K.}~\bibnamefont
  {Zhou}}, \bibinfo {author} {\bibfnamefont {G.}~\bibnamefont {Salomon}}, \
  and\ \bibinfo {author} {\bibfnamefont {E.}~\bibnamefont {Wu}},\ }\href@noop
  {} {\bibfield  {journal} {\bibinfo  {journal} {Int. J. Robust and Nonlin.
  Contr.}\ }\textbf {\bibinfo {volume} {9}},\ \bibinfo {pages} {183} (\bibinfo
  {year} {1999})}\BibitemShut {NoStop}%
\bibitem [{\citenamefont {Farrell}\ and\ \citenamefont
  {Ioannou}(2001)}]{FaIo01}%
  \BibitemOpen
  \bibfield  {author} {\bibinfo {author} {\bibfnamefont {B.~F.}\ \bibnamefont
  {Farrell}}\ and\ \bibinfo {author} {\bibfnamefont {P.~J.}\ \bibnamefont
  {Ioannou}},\ }\href@noop {} {\bibfield  {journal} {\bibinfo  {journal} {J.
  Atmos. Sci.}\ }\textbf {\bibinfo {volume} {58}},\ \bibinfo {pages} {2771}
  (\bibinfo {year} {2001})}\BibitemShut {NoStop}%
\bibitem [{\citenamefont {Varga}(2001)}]{Varga01}%
  \BibitemOpen
  \bibfield  {author} {\bibinfo {author} {\bibfnamefont {A.}~\bibnamefont
  {Varga}},\ }in\ \href@noop {} {\emph {\bibinfo {booktitle} {Proc. of IFAC
  Workshop on Periodic Control Systems, Como, Italy}}}\ (\bibinfo {year}
  {2001})\ pp.\ \bibinfo {pages} {177--182}\BibitemShut {NoStop}%
\bibitem [{\citenamefont {Crutchfield}\ and\ \citenamefont
  {Packard}(1983)}]{CruPack83}%
  \BibitemOpen
  \bibfield  {author} {\bibinfo {author} {\bibfnamefont {J.~P.}\ \bibnamefont
  {Crutchfield}}\ and\ \bibinfo {author} {\bibfnamefont {N.~H.}\ \bibnamefont
  {Packard}},\ }\href@noop {} {\bibfield  {journal} {\bibinfo  {journal}
  {Physica D}\ }\textbf {\bibinfo {volume} {7}},\ \bibinfo {pages} {201}
  (\bibinfo {year} {1983})}\BibitemShut {NoStop}%
\bibitem [{\citenamefont {{Daw}}\ \emph {et~al.}(2003)\citenamefont {{Daw}},
  \citenamefont {{Finney}},\ and\ \citenamefont {{Tracy}}}]{DFT03}%
  \BibitemOpen
  \bibfield  {author} {\bibinfo {author} {\bibfnamefont {C.~S.}\ \bibnamefont
  {{Daw}}}, \bibinfo {author} {\bibfnamefont {C.~E.~A.}\ \bibnamefont
  {{Finney}}}, \ and\ \bibinfo {author} {\bibfnamefont {E.~R.}\ \bibnamefont
  {{Tracy}}},\ }\href {\doibase 10.1063/1.1531823} {\bibfield  {journal}
  {\bibinfo  {journal} {Rev. Sci. Instrum.}\ }\textbf {\bibinfo {volume}
  {74}},\ \bibinfo {pages} {915} (\bibinfo {year} {2003})}\BibitemShut
  {NoStop}%
\bibitem [{\citenamefont {{Buhl}}\ and\ \citenamefont
  {{Kennel}}(2005)}]{BuKe05}%
  \BibitemOpen
  \bibfield  {author} {\bibinfo {author} {\bibfnamefont {M.}~\bibnamefont
  {{Buhl}}}\ and\ \bibinfo {author} {\bibfnamefont {M.~B.}\ \bibnamefont
  {{Kennel}}},\ }\href@noop {} {\bibfield  {journal} {\bibinfo  {journal}
  {Phys. Rev. E}\ }\textbf {\bibinfo {volume} {71}},\ \bibinfo {pages} {046213}
  (\bibinfo {year} {2005})}\BibitemShut {NoStop}%
\bibitem [{\citenamefont {Ulam}(1960)}]{Ulam60}%
  \BibitemOpen
  \bibfield  {author} {\bibinfo {author} {\bibfnamefont {S.~M.}\ \bibnamefont
  {Ulam}},\ }\href@noop {} {\emph {\bibinfo {title} {A Collection of
  Mathematical Problems}}}\ (\bibinfo  {publisher} {Interscience Publishers},\
  \bibinfo {address} {New York},\ \bibinfo {year} {1960})\BibitemShut {NoStop}%
\bibitem [{\citenamefont {Froyland}(2001)}]{Froy-2}%
  \BibitemOpen
  \bibfield  {author} {\bibinfo {author} {\bibfnamefont {G.}~\bibnamefont
  {Froyland}},\ }in\ \href@noop {} {\emph {\bibinfo {booktitle} {Nonlinear
  Dynamics and Statistics: Proc. Newton Inst., Cambridge 1998}}},\ \bibinfo
  {editor} {edited by\ \bibinfo {editor} {\bibfnamefont {A.}~\bibnamefont
  {Mees}}}\ (\bibinfo  {publisher} {Birkh\"{a}user},\ \bibinfo {address}
  {Boston},\ \bibinfo {year} {2001})\ pp.\ \bibinfo {pages}
  {281--321}\BibitemShut {NoStop}%
\bibitem [{\citenamefont {Chappell}\ \emph {et~al.}(2013)\citenamefont
  {Chappell}, \citenamefont {Tanner}, \citenamefont {L\"ochel},\ and\
  \citenamefont {S{\o}ndergaard}}]{Tanner_cars}%
  \BibitemOpen
  \bibfield  {author} {\bibinfo {author} {\bibfnamefont {D.~J.}\ \bibnamefont
  {Chappell}}, \bibinfo {author} {\bibfnamefont {G.}~\bibnamefont {Tanner}},
  \bibinfo {author} {\bibfnamefont {D.}~\bibnamefont {L\"ochel}}, \ and\
  \bibinfo {author} {\bibfnamefont {N.}~\bibnamefont {S{\o}ndergaard}},\ }\href
  {\doibase 10.1098/rspa.2013.0153} {\bibfield  {journal} {\bibinfo  {journal}
  {Proc. R. Soc. A}\ }\textbf {\bibinfo {volume} {469}},\ \bibinfo {pages}
  {20130153} (\bibinfo {year} {2013})}\BibitemShut {NoStop}%
\bibitem [{\citenamefont {Sinai}(1972)}]{sinai}%
  \BibitemOpen
  \bibfield  {author} {\bibinfo {author} {\bibfnamefont {Y.~G.}\ \bibnamefont
  {Sinai}},\ }\href@noop {} {\bibfield  {journal} {\bibinfo  {journal} {Russian
  Math. Surveys}\ }\textbf {\bibinfo {volume} {166}},\ \bibinfo {pages} {21}
  (\bibinfo {year} {1972})}\BibitemShut {NoStop}%
\bibitem [{\citenamefont {Bowen}(1975)}]{bowen}%
  \BibitemOpen
  \bibfield  {author} {\bibinfo {author} {\bibfnamefont {R.}~\bibnamefont
  {Bowen}},\ }\href@noop {} {\emph {\bibinfo {title} {Equilibrium States and
  the Ergodic Theory of {A}nosov Diffeomorphisms}}}\ (\bibinfo  {publisher}
  {Springer},\ \bibinfo {address} {Berlin},\ \bibinfo {year}
  {1975})\BibitemShut {NoStop}%
\bibitem [{\citenamefont {Ruelle}(1978)}]{ruelle}%
  \BibitemOpen
  \bibfield  {author} {\bibinfo {author} {\bibfnamefont {D.}~\bibnamefont
  {Ruelle}},\ }\href@noop {} {\emph {\bibinfo {title} {Statistical Mechanics,
  Thermodynamic Formalism}}}\ (\bibinfo  {publisher} {Addison-Wesley},\
  \bibinfo {address} {Reading, MA},\ \bibinfo {year} {1978})\BibitemShut
  {NoStop}%
\bibitem [{\citenamefont {Lozi}(1978)}]{lozi2}%
  \BibitemOpen
  \bibfield  {author} {\bibinfo {author} {\bibfnamefont {R.}~\bibnamefont
  {Lozi}},\ }\href {\doibase 10.1051/jphyscol:1978505} {\bibfield  {journal}
  {\bibinfo  {journal} {J. Phys. (Paris) Colloq.}\ }\textbf {\bibinfo {volume}
  {39}},\ \bibinfo {pages} {C5} (\bibinfo {year} {1978})}\BibitemShut {NoStop}%
\bibitem [{\citenamefont {Gaji\'c}\ and\ \citenamefont
  {Qureshi}(1995)}]{GaQu95}%
  \BibitemOpen
  \bibfield  {author} {\bibinfo {author} {\bibfnamefont {Z.}~\bibnamefont
  {Gaji\'c}}\ and\ \bibinfo {author} {\bibfnamefont {M.}~\bibnamefont
  {Qureshi}},\ }\href@noop {} {\emph {\bibinfo {title} {Lyapunov Matrix
  Equation in System Stability and Control}}}\ (\bibinfo  {publisher} {Academic
  Press},\ \bibinfo {address} {New York},\ \bibinfo {year} {1995})\BibitemShut
  {NoStop}%
\bibitem [{\citenamefont {Devaney}(1989)}]{devnmap}%
  \BibitemOpen
  \bibfield  {author} {\bibinfo {author} {\bibfnamefont {R.~L.}\ \bibnamefont
  {Devaney}},\ }\href@noop {} {\emph {\bibinfo {title} {An Introduction to
  Chaotic Dynamical systems}}}\ (\bibinfo  {publisher} {Addison-Wesley},\
  \bibinfo {address} {Red-wood City},\ \bibinfo {year} {1989})\BibitemShut
  {NoStop}%
\bibitem [{\citenamefont {Gottfried}\ and\ \citenamefont
  {Yan}(2003)}]{GottYan}%
  \BibitemOpen
  \bibfield  {author} {\bibinfo {author} {\bibfnamefont {K.}~\bibnamefont
  {Gottfried}}\ and\ \bibinfo {author} {\bibfnamefont {T.}~\bibnamefont
  {Yan}},\ }\enquote {\bibinfo {title} {Quantum mechanics: Fundamentals},}\ \
  (\bibinfo  {publisher} {Springer},\ \bibinfo {address} {New York},\ \bibinfo
  {year} {2003})\ Chap.\ \bibinfo {chapter} {Low-Dimensional Systems}, pp.\
  \bibinfo {pages} {181--184},\ \bibinfo {edition} {2nd}\ ed.\BibitemShut
  {Stop}%
\bibitem [{\citenamefont {Ermann}\ and\ \citenamefont
  {Shepelyansky}(2012)}]{ErmShep12}%
  \BibitemOpen
  \bibfield  {author} {\bibinfo {author} {\bibfnamefont {L.}~\bibnamefont
  {Ermann}}\ and\ \bibinfo {author} {\bibfnamefont {D.}~\bibnamefont
  {Shepelyansky}},\ }\href@noop {} {\bibfield  {journal} {\bibinfo  {journal}
  {Eur. Phys. J. B}\ }\textbf {\bibinfo {volume} {75}},\ \bibinfo {pages} {299}
  (\bibinfo {year} {2012})}\BibitemShut {NoStop}%
\bibitem [{\citenamefont {Cvitanovi{\'c}}(2013)}]{focusPOT}%
  \BibitemOpen
  \bibfield  {author} {\bibinfo {author} {\bibfnamefont {P.}~\bibnamefont
  {Cvitanovi{\'c}}},\ }\href {\doibase 10.1017/jfm.2013.198} {\bibfield
  {journal} {\bibinfo  {journal} {J. Fluid Mech. Focus on Fluids}\ }\textbf
  {\bibinfo {volume} {726}},\ \bibinfo {pages} {1} (\bibinfo {year}
  {2013})}\BibitemShut {NoStop}%
\bibitem [{\citenamefont {Nagata}(1990)}]{N90}%
  \BibitemOpen
  \bibfield  {author} {\bibinfo {author} {\bibfnamefont {M.}~\bibnamefont
  {Nagata}},\ }\href@noop {} {\bibfield  {journal} {\bibinfo  {journal} {J.
  Fluid Mech.}\ }\textbf {\bibinfo {volume} {217}},\ \bibinfo {pages} {519}
  (\bibinfo {year} {1990})}\BibitemShut {NoStop}%
\bibitem [{\citenamefont {Gibson}\ \emph {et~al.}(2008)\citenamefont {Gibson},
  \citenamefont {Halcrow},\ and\ \citenamefont {Cvitanovi{\'c}}}]{GHCW07}%
  \BibitemOpen
  \bibfield  {author} {\bibinfo {author} {\bibfnamefont {J.~F.}\ \bibnamefont
  {Gibson}}, \bibinfo {author} {\bibfnamefont {J.}~\bibnamefont {Halcrow}}, \
  and\ \bibinfo {author} {\bibfnamefont {P.}~\bibnamefont {Cvitanovi{\'c}}},\
  }\href@noop {} {\bibfield  {journal} {\bibinfo  {journal} {J. Fluid Mech.}\
  }\textbf {\bibinfo {volume} {611}},\ \bibinfo {pages} {107} (\bibinfo {year}
  {2008})},\ \bibinfo {note} {\arXiv{0705.3957}}\BibitemShut {NoStop}%
\bibitem [{\citenamefont {Gaspard}(2002)}]{gasp02}%
  \BibitemOpen
  \bibfield  {author} {\bibinfo {author} {\bibfnamefont {P.}~\bibnamefont
  {Gaspard}},\ }\href {\doibase 10.1023/A:1013167928166} {\bibfield  {journal}
  {\bibinfo  {journal} {J. Stat. Phys.}\ }\textbf {\bibinfo {volume} {106}},\
  \bibinfo {pages} {57} (\bibinfo {year} {2002})}\BibitemShut {NoStop}%
\end{thebibliography}%

\ifboyscout
\newpage
\input{../blog/notesHenCviLip}
\input{../blog/flotHenCviLip}
\fi

\end{document}